\documentclass[reprint, amsmath,amssymb,nofootinbib,floatfix, aps, longbibliography]{revtex4-1}
\usepackage{graphicx}
\usepackage{dcolumn}
\usepackage{bm}
\usepackage[colorlinks=true,urlcolor=rossos,anchorcolor=black,citecolor=mygreen,filecolor=black,linkcolor=black,menucolor=blue,linktocpage=true]{hyperref} % should be commented out if the tex file will be compiled with latex in arXiv!!! (pdflatex is fine) 
\usepackage{booktabs}
\usepackage{colortbl}
\usepackage{collcell}
\usepackage{enumerate}
\usepackage{float}
\usepackage{verbatim}
\usepackage{multirow}
\usepackage{xparse}

\definecolor{rossos}{cmyk}{0,1,1,0.55}
\definecolor{mygreen}{rgb}{0.27, 0.64, 0.48}
\definecolor{mygray}{gray}{.95}

\newcommand{\virg}[1]{``#1''}

\def\min{\text{min}}
\def\max{\text{max}}

\def\DM{\text{DM}}

\def\LOS{\text{LOS}}
\def\BH{\text{BH}}
\def\doppler{\mathcal{D}}

\begin{document}

\title{Blazar-Boosted Dark Matter: Novel Signatures via Elastic and Inelastic Scattering}

\author{Jin-Wei Wang}
\email{jinwei.wang@uestc.edu.cn}
\affiliation{School of Physics, University of Electronic Science and Technology of China, Chengdu 611731, China}

\begin{abstract}
Blazar-Boosted Dark Matter (BBDM) is a novel mechanism whereby dark matter (DM) particles are accelerated to ultrarelativistic energies through interactions with blazar jets. Focusing on a vector portal DM model, we systematically investigate both elastic and inelastic scattering processes between DM and protons. By analyzing multi-messenger data from ground- and space-based observatories, we derive stringent constraints on the DM-proton scattering cross section $\sigma_{\chi p}$. Our results improve upon previous limits from the constant cross section and pure elastic scattering scenarios by several orders of magnitude.
The distinctive high-energy gamma ray and neutrino fluxes produced through deep inelastic scattering provide powerful signatures for BBDM indirect detection, enabling constraints that significantly surpass those from traditional direct detection experiments.
Notably, our results suggest that DM could be a new source of high-energy neutrinos from blazars, potentially offering an explanation for IceCube's observation of TXS 0506+056.
\end{abstract}

\maketitle

\section{Introduction}
\label{sec:intro}
The existence of dark matter (DM) is robustly supported by diverse astrophysical and cosmological observations, including galaxy rotation curves, the dynamics of galaxy clusters, and anisotropies in the cosmic microwave background (CMB) \cite{hep-ph/0404175,Taoso:2007qk,Bertone:2016nfn,1807.06209,Arbey:2021gdg,Cirelli:2024ssz}. Despite this compelling evidence, the particle nature of DM remains elusive, representing one of the most profound and unresolved questions at the forefront of particle physics.
Assuming that DM interacts extremely weakly with ordinary matter, direct detection experiments such as LUX-ZEPLIN (LZ) \cite{LZ:2015kxe,LZ:2018qzl,LZ:2022lsv}, XENONnT \cite{XENON:2022ltv,XENON:2023cxc}, and PandaX-4T \cite{PandaX:2018wtu,PandaX-4T:2021bab} hold significant promise for observing DM scattering events with target nuclei. However, for DM masses below $\mathcal{O}(1)$ GeV, the typical kinetic energy of DM particles in the Milky Way halo is insufficient to trigger a detectable signal at an experimental threshold of $\mathcal{O}(1)$ keV, causing a sharp decrease in the sensitivity to candidates with low-mass DM.

To overcome this limitation, several scenarios have been proposed that involve boosted DM populations \cite{1405.7370,1506.04316,1708.03642,1709.06573,PhysRevLett.122.171801,Yin:2018yjn,Elor:2021swj}. For example, Ref.\cite{PhysRevLett.122.171801} introduces an innovative mechanism in which DM particles in the Milky Way halo are accelerated through elastic collisions with high-energy galactic cosmic rays (CRs). This small yet inevitable DM component, referred to as cosmic-ray-boosted DM (CRDM), carries sufficient energy to impose constraints on the DM-proton scattering cross section $\sigma_{\chi p}$ for sub-GeV DM. 
This boosted DM component offers a new avenue for the detection of light DM using conventional direct detection techniques. 

Recently,~\cite{wang2021direct,Granelli:2022ysi} proposed a novel DM acceleration mechanism associated with blazars, termed Blazar-Boosted Dark Matter (BBDM). Blazars are a type of active galactic nuclei (AGN) with a pair of extremely relativistic back-to-back jets, one of them in close alignment to our line of sight (LOS). In addition, it is believed that at the centers of blazars reside supermassive black holes (BH), which, through adiabatic growth, could lead to the formation of regions with significantly enhanced DM densities known as DM spikes \cite{Gondolo:1999ef,Ullio:2001fb,Merritt:2003qk,Gnedin:2003rj}. Through the scatterings with high-energy particles in blazar jets, DM particles can be accelerated to high velocities.  Refs.~\cite{wang2021direct,Granelli:2022ysi} have shown that blazars are more efficient DM accelerators and can generate much higher boosted DM flux than that from galactic cosmic rays.

%In Ref.~\cite{wang2021direct,Granelli:2022ysi} the authors focused on the blazars TXS 0506+056 and BL Lacertae and derived the corresponding constraints on $\sigma_{\chi p}$ from the results of direct DM detectors (e.g.~XENON1T \cite{XENON:2017lvq}), as well as neutrino detectors (e.g.~MiniBooNE \cite{MiniBooNE:2008paa} and Borexino \cite{Borexino:2000uvj}). 

For simplicity, the authors of Refs.~\cite{wang2021direct,Granelli:2022ysi} assume a constant cross section for DM interactions with protons ($\sigma_{\chi p}$) and electrons ($\sigma_{\chi e}$), and only consider the elastic scattering process. Here we argue that these simplified treatments are not enough to give reliable results, and some new phenomenologies may even disappear due to these simplifications. For example, it is known that the blazar jet can pose extremely high luminosity, says $\sim 10^{48}$ erg/s, and the energy spectrum of jet particles is very wide, e.g., from GeV to above PeV \cite{Boettcher:2013wxa,Cerruti:2018tmc} (see Ref. \cite{Cerruti:2020lfj} for a recent review of blazar jet model). In this case, it is crucial to consider the energy dependence of the scattering cross section, as in practical and specific models, the scattering cross section is almost invariably energy-dependent. Moreover, given the extremely high energies of particles within blazar jets, inelastic scattering processes become significant and must be taken into account. This would not only affect the BBDM flux but could also provide new avenues for detecting DM.

Specifically,  when protons collide with DM at center-of-mass energy $\gtrsim$ GeV, nucleons can undergo excitation, leading to the production of secondary gamma ray and neutrinos in the decay sequence \cite{Cyburt:2002uw,Hooper:2018bfw}. At even higher energy scales, DM can interact with individual quarks within nucleons, resulting in deep inelastic scatterings (DIS). These DIS processes will also generate secondary neutrinos and gamma ray from the hadronization and the subsequent decay of mesons (e.g., $\pi^\pm$ and $\pi^0$), which means the DM within blazars could serve as the additional new source of high-energy gamma ray and neutrinos.
Remarkably, observational evidence supports the astrophysical relevance of such processes. In 2017, the IceCube Neutrino Observatory reported a significant flux of high-energy neutrinos spatially and temporally coincident with the blazar TXS 0506+056 \cite{IceCube:2018dnn,IceCube:2018cha,Padovani:2018acg}. While conventional blazar models attribute this signal to accelerated protons in relativistic jets, a compelling alternative interpretation posits that these neutrinos could originate from DIS interactions between DM and Standard Model (SM) particles in the blazar environment. 
This duality highlights the dual role of secondary emissions: they serve both as probes of DM properties and as diagnostics of blazar astrophysics (e.g., jet dynamics and magnetic field configurations \cite{Cerruti:2020lfj}), which is fundamentally different from considering only the elastic scattering process~\cite{wang2021direct,Granelli:2022ysi,Herrera:2023nww}. 

As an extension of previous work~\cite{wang2021direct,Granelli:2022ysi}, in this paper we study the both elastic and inelastic effects of BBDM by considering a specific vector portal fermionic DM model. The remainder of this work is organized as follows.
In Sec.~\ref{sec:model}, we give a brief introduction to the model setup and present calculations of the elastic and inelastic cross section between DM and protons. In Sec.~\ref{sec:BBDM}, we analyze the DM profile in blazars and derive the resulting BBDM flux. In Sec.~\ref{sec:gammav}, we calculate the high-energy neutrino and gamma ray flux induced by the DIS process.  The computation of the constraints on $\sigma_{\chi p}$ from both ``direct" and  ``indirect"  experiments are summarised in Sec.~\ref{sec:constraints}. Conclusions are presented in Sec.~\ref{sec:Conclusion}.

\section{elastic and inelastic cross section between DM and protons}
\label{sec:model}
In this work, we consider a benchmark model where the new sector consists of a fermionic DM $\chi$ and a massive vector mediator $V_\mu$. Extending to other models, such as scalar mediator models, is straightforward \cite{DeMarchi:2024riu,Su:2022wpj}. The corresponding Lagrangian can be expressed as \cite{Guo:2020oum,Su:2023zgr}
\begin{equation}
    \begin{split}
    \mathcal{L} &\supset \bar{\chi}(i\partial_\mu \gamma^\mu - m_\chi)\chi + g_\chi \bar{\chi}\gamma^\mu \chi V_\mu\\
    &+ \sum_q g_q \bar{q}\gamma^\mu q V_\mu + \frac{1}{2} m_V^2 V_\mu V^\mu
    \end{split}
\end{equation}
where $m_\chi$ and $m_V$ are the DM and vector mediator mass, $g_\chi$ and $g_q$ are the coupling constant between vector mediator and DM and quarks, respectively. Note that we have assumed a universal coupling constant between the vector mediator and quarks of different flavors.
The differential cross section for elastic scattering between DM and protons is given by \cite{Guo:2020oum,Su:2023zgr}
\begin{align}
 \frac{d\sigma_{\rm el}}{dQ^2} = \frac{ g_p^2g_\chi^2}{4\pi \beta^2} \frac{G^2_p(Q^2)}{(Q^2+m_V^2)^2} \left[ 1 -   \frac{\beta^2 Q^2}{Q^2_{\rm max}} + \frac{Q^4}{8 m_p^2 E_\chi^2} \right] \;,  \label{eq:dxsec_elas}  
\end{align}
where $m_p$ is proton mass,  $g_p = 3 g_q$ is the coupling constant between DM and proton, $Q^2 \equiv -q^2$ is the positive four-momentum transfer squared, $\beta=|{\bf p}_\chi|/E_\chi$ is the incoming velocity of DM, and $G_p$ is the form factor of the proton, which takes a dipole form,
\begin{align}
G_p(Q^2) = \frac{1}{\big(1+Q^2/\Lambda_p^2\big)^2}\;,  
\label{eq:form}
\end{align}
with $\Lambda_p \approx$ 770 MeV \cite{Angeli:2004kvy}. 
The maximal momentum transfer squared, $Q^2_{\rm max}$, is 
\begin{align}
 Q^2_{\rm max} = \frac{4(E_\chi^2-m_\chi^2)m_p^2}{m_\chi^2 + m_p^2 + 2 m_p E_\chi}\;, \label{eq:Qmax}
\end{align}

For DIS process, we adopt the commonly used Bjorken scaling variables \cite{Buckley:2014ana,AbdulKhalek:2022fyi}. For the sake of simplicity, we omit the detailed calculation process, and only give the key calculation results, with the detailed calculations available in Refs. \cite{ParticleDataGroup:2024cfk,Berger:2018urf,Su:2023zgr}. The differential cross section is given by 
\begin{align}
\frac{d^2\sigma_{\rm DIS}}{d\nu dQ^2} = \frac{1}{2m_p  E_\chi^2 y} \frac{d^2\sigma_{\rm DIS}}{dxdy}\;,  
\label{eq:ddxsec_dis}
\end{align}
where $\nu$ is the energy transfer of DM in the proton rest
frame, $x$ and $y$ are the Bjorken variables,
\begin{equation}
    \begin{split}
&\frac{d^2\sigma_{\rm DIS}}{dxdy} = \frac{g_\chi^2 g_q^2}{4\pi(Q^2+m_V^2)^2} \frac{2 E_\chi^3 m_p}{E_\chi^2-m_\chi^2}\\
&\times\left(1-y+\frac{y^2}{2} - \frac{xy m_p}{2 E_\chi} -\frac{m_\chi^2}{2 E_\chi m_p} \frac{y}{x}\right)F_2(x, Q^2)\;,    
\end{split}
\end{equation}
and the corresponding structure function is    
\begin{align}
F_2(x,Q^2) = \sum_{f=u,d,c,s,b} \Big[f(x,Q^2) + \bar f(x,Q^2)\Big] x,       
\end{align}
where $f$ $(\bar f)$ the parton distribution functions (PDFs) of quarks (anti-quarks). In our study, we use the MSTW 2008 NNLO PDFs \cite{Martin:2009iq}. By integrating out the
energy transfer $\nu$ we can obtain
\begin{equation}
    \frac{d\sigma_\text{DIS}}{dQ^2}=\int \frac{d^2\sigma_{\rm DIS}} {d\nu dQ^2} d\nu.
    \label{eq:disQ2}
\end{equation}
Note that the above results for elastic and inelastic scattering are obtained in the proton rest frame, while in the observer frame the initial DM is at rest. The transformation between different reference frames is straightforward and intuitive. By using $T_\chi = Q^2/2m_\chi$, where $T_\chi$ is the DM kinetic energy after scattering, we can get
\begin{equation}
    \frac{d\sigma_{\chi p}}{dT_\chi}=2m_\chi \left(\frac{d\sigma_{\rm el}}{dQ^2} +  \frac{d\sigma_{\rm DIS}}{dQ^2}\right).
\end{equation}

It is clearly shown that the differential cross section between DM and protons depends on $m_V$ and $m_\chi$. In this work, we are more interested in the light DM scenario, says $m_\chi \lesssim 0.1$ GeV, which is more attractive for boosted DM models. While for $m_V$, we consider two benchmark cases: $m_V = 1$ GeV and $m_V=10$ GeV, and the influence of other values of $m_V$ can be inferred from benchmark results.

\section{Blazar-Boosted Dark Matter Flux}
\label{sec:BBDM}
Knowing the DM-proton cross section (see Sec. \ref{sec:model}), it is crucial to know the blazar jet spectrum and DM profile to calculate the flux of BBDM. Following previous works \cite{wang2021direct,Granelli:2022ysi}, we also adopt the \virg{blob geometry} to describe the blazar jet \cite{Dermer2009-DERHER}, where  protons move isotropically in the blob frame with a power-law energy distribution,
and, in the BH center-of-mass rest frame (also observer's frame), the blob itself moves along the jet axis with speed $\beta_B$, and its corresponding Lorentz boost factor reads $\Gamma_B \equiv (1-\beta_B^2)^{-1/2}$.
The misalignment angle between the jet axis and the observer's line-of-sight is $\theta_\LOS$. 
The desired jet spectrum in the observer's frame can be derived from a Lorentz boost transformation and can be expressed as(see Ref. \cite{wang2021direct,Granelli:2022ysi} for a detailed derivation)
\begin{equation}\label{eq:CRSpectrum}
\begin{split}
    \frac{d\Gamma_p}{dT_pd\Omega} 
    =&\, \frac{1}{4\pi}c_p\,\left(1+\frac{T_p}{m_p}\right)^{-\alpha_p}\\ &\times\frac{\beta_p(1-\beta_p\beta_B  \mu)^{-\alpha_p} \Gamma_B^{-\alpha_p}}{\sqrt{(1-\beta_p \beta_B \mu)^2 - (1-\beta_p^2)(1-\beta_B^2)}}\,,
    \end{split}
\end{equation}
where $\alpha_j$ is the spectral power index, $T_{p}$ and $\beta_p = \left[1-m_p^2/(T_p+m_p)^2\right]^{1/2}$ are respectively the kinetic energy and speed of the particle, $c_p$ is the normalisation constant that can be computed from the luminosity $L_p$ (see, e.g., \cite{PhysRevD.82.083514}) and, finally, $\mu$ is the cosine of the angle between the particle's direction of motion and the jet axis. In this work, we focus on two blazars, i.e., TXS 0506+056 and BL Lacertae as in Refs. \cite{wang2021direct,Granelli:2022ysi}, and the detailed parameters are displayed in TABLE. \ref{Tab:HadronicModel}.

As demonstrated in Ref.~\cite{PhysRevLett.83.1719}, the adiabatic growth of a BH at the center of a DM halo gravitationally focuses DM particles, forming an ultra-dense spike. For an initial power-law profile, i.e.~$\rho(r) \propto r^{-\gamma}$, the post-evolution density profile follows $\rho'(r)  \propto  r^{-(9-2\gamma)/(4-\gamma)}$, called the spike distribution \cite{PhysRevLett.83.1719}. In our analysis, we adopt $\gamma=1$, which corresponds to an initial Navarro-Frenk-White profile \cite{NFW_profile}. Note that this idealized spike structure may be altered by DM annihilation: self-annihilation processes can deplete the DM density in the innermost region, converting the steep 
cusp into a flattened core. The previous work considered several typical values of DM annihilation cross section, here, with the specific model, we can have a more proper evaluation. 

\begin{table}
\centering
\begin{ruledtabular}
\begin{tabular}{ccc}
    \rowcolor[gray]{.95}
    \multicolumn{3}{c}{\bf (Lepto-)Hadronic Model Parameters}\\
    \colrule
     \rule{0pt}{2.5ex} 
    ~~~Parameter (unit)~~~ & TXS 0506+056~~~ & BL Lacertae~~~ \\
    \colrule
    \rule{0pt}{2.5ex} 
     $z$   &0.337 & 0.069 \\
     $d_L$ (Mpc)    & 1835.4 & 322.7 \\
     $M_\BH$ ($M_\odot$) &$3.09\times10^{8} $ &$8.65\times 10^7$\\
    $\mathcal{D}$  &40$^\star$& 15  \\
    $\Gamma_B$ &20& 15  \\
     $\theta_\LOS\, (^\circ)$  &$0$&$3.82$  \\
    $\alpha_p$   & $2.0^\star$& $2.4$\\
 %    $\alpha_e$   & $2.0^\star$& $3.5$\\
    $\gamma'_{\min,\,p}$ & 1.0 & 1.0  \\
     $\gamma'_{\max,\,p}$ & $5.5\times10^{7^\star}$& $1.9\times10^9$  \\
 %    $\gamma'_{\min,\,e}$ & 500 & 700  \\
 %    $\gamma'_{\max,\,e}$ & $1.3\times10^{4^\star}$& $1.5\times10^4$  \\
     $L_p$ (erg/s)  & $2.55 \times 10^{48^\star}$& $9.8 \times 10^{48}$  \\
%     $L_e$ (erg/s)  & $1.32 \times 10^{44^\star}$& $8.7 \times 10^{42}$  \\
    \bottomrule
\end{tabular}
\end{ruledtabular}
\caption{The model parameters for the blazars TXS 0506+056 (Lepto-Hadronic) \cite{TXS, TXS_2} and BL Lacertae (Hadronic) \cite{1304.0605} used in our calculations. 
The quantities flagged with a star ($^\star$) correspond to mean values computed from the ranges given in the second column of Table 1 of Ref.~\cite{TXS_2}.
In the model fitting, the assumption of $\doppler = 2\Gamma_B$ ($\Gamma_B$) is used for TXS 0506+056 (BL Lacertae). The redshift $z$ \cite{BLRedshift, 1802.01939}, luminosity distance $d_L$, and BH mass $M_\BH$ \cite{Titarchuk:2017jwu, 1901.06998} for the two considered sources are also reported.}
\label{Tab:HadronicModel}
\end{table}

For \( m_\chi \lesssim 0.1 \) GeV, the annihilation channel into two mesons (e.g., \(\pi^+ \pi^-\)) is kinematically forbidden. However, alternative annihilation channels remain viable depending on the DM mass. For instance, if \( m_\chi > m_e \), DM can annihilate into electron-positron pairs (\( e^+ e^- \)) and/or into photons through loop processes \cite{Diamond:2023fsm}. Due to the small mass of \( m_\chi \) and the limited initial kinematic energy, the annihilation into electron-positron pairs occurs predominantly via a proton loop, as illustrated in Fig. 1(b) of Ref. \cite{Diamond:2023fsm}. Notably, contributions from mesons such as pions and kaons are absent in our model due to the universal choice of the coupling constant \( g_q \). 
For \( m_\chi < m_e \), DM annihilation into three photons (\(\gamma \gamma \gamma\)) becomes dominant, while the two-photon final state (\(\gamma \gamma\)) is forbidden by Furry's theorem \cite{Furry:1937zz}. The annihilation cross section into electron-positron pairs can be expressed as
\begin{equation}
    \sigma v_{e^+e^-} \simeq \frac{\alpha^2 g_\chi^2 m_\chi^2}{144 \pi^3 m_V^4}\left(4 g_p \text{Log}\left[\frac{4\pi e^{-\gamma_E}\mu^2}{m_p^2}\right]\right)^2.
\end{equation}
Here we set $\mu=m_p$ as a cutoff scale for low energy ChEFT with Baryons \cite{Becher:1999he}, $\gamma_E = 0.577$ is the Euler Mascheroni constant. The annihilation cross section to $\gamma \gamma \gamma$ can be expressed as
\begin{equation}
    \sigma v_{\gamma \gamma \gamma} \simeq \frac{\alpha^3 g_\chi^2 g_p^2 m_\chi^2}{16 \pi^4 m_V^4}.
\end{equation}
By adopting a set of benchmark values: $m_\chi=0.01$ GeV, $m_V=1$ GeV, $g_\chi =g_q=0.01$. The corresping $\sigma v_{e^+e^-} \simeq 2.56 \times 10^{-45} ~\text{cm}^{2}$ and $\sigma v_{\gamma \gamma \gamma} \simeq 8.77 \times 10^{-49} ~\text{cm}^{2}$. 
\textcolor{black}{We have verified that for $\sigma v \lesssim 10^{-40} \, \text{cm}^2$, DM annihilation does not significantly affect the DM spike profile in the central region of the blazar (see Fig. 1 in Ref. \cite{wang2021direct}).}
In this case, we can ignore the DM annihilation effects, and adopt a spike DM profile.  We adopt the same normalization condition for $\rho'(r)$ as in Ref.~\cite{wang2021direct,Ullio:2001fb}, that is
\begin{equation}\label{eq:DM_condition}
\int_{4R_S}^{10^5 R_S} 4\pi r^2 \rho'(r) dr \simeq M_\BH\,,
\end{equation}
where $R_S$ is the Schwarzschild radius of the central BH. 
\textcolor{black}{DM particles on smaller orbits ($< 4 R_s$) are accreted onto the supermassive BH \cite{Gondolo:1999ef}, while the upper limit of $10^5 R_s$ is pertinent for determining BH masses \cite{Neumayer_2010,Gebhardt_2009}.}
What is relevant to derive the flux of BBDM is the DM line-of-sight integral $\Sigma_\DM$, which is defined as
\begin{equation}\label{eq:deltaDM}
    \Sigma_\DM(r) \equiv \int_{4 R_S}^{r}  \rho_\DM(r')\,dr'.
\end{equation}
Given that $\Sigma_\DM (r)$ tends to a constant value for $r\gg 10$ pc \cite{wang2021direct}, we can factor the effects of the DM profile into $\Sigma_\DM^\text{tot}\equiv \Sigma_\DM(r\simeq 100\,\text{pc})$. 
\textcolor{black}{Note that the influence of varying the lower integration limit on $\Sigma_\DM$ has been thoroughly discussed in Ref.~\cite{wang2021direct}.}
With the above jets spectrum and DM profile, the BBDM flux at Earth can be expressed as\footnote{\textcolor{black}{We note that the $\nu$-dependence in the DIS case implicitly includes a dependence on the scattering angle $\theta$, which is not fixed by kinematics in inelastic scattering. Therefore, Eq. \eqref{eq:spectrumDM} should be considered an approximation for DIS. However, we have found that in the mass range $m_\chi \lesssim 100~\mathrm{MeV}$, which is the main focus of this work, the difference between this approximation and the exact result becomes negligible as the $m_\chi$ decreases. For instance, at $m_\chi = 10~\mathrm{MeV}$, the two results are nearly identical. Since the high-energy neutrino and photon signatures from DIS impose the most stringent constraints (see Fig.~\ref{fig:IDexp}), we consider this approximation sufficient for our purposes. We are grateful to the anonymous Referee for raising this point and to Alessandro Granelli for helpful discussions.}} \cite{wang2021direct}
\begin{equation}
\label{eq:spectrumDM}
    \frac{d\Phi_\chi}{dT_\chi} =\frac{\Sigma_\DM^\text{tot}}{2\pi m_\chi d_L^2}\int_{0}^{2\pi}\,d\phi_s\int_{T_j^\min(T_\chi,\phi_s)}^{T_j^\max(T_\chi,\phi_s)}\frac{d\sigma_{\chi p}}{dT_\chi}\frac{d\Gamma_j}{dT_j d\Omega}dT_j\,,
\end{equation}
where $\phi_s$ is the azimuth angle with respect to the line-of-sight.

\begin{figure}
\centering
\includegraphics[width=0.45\textwidth]{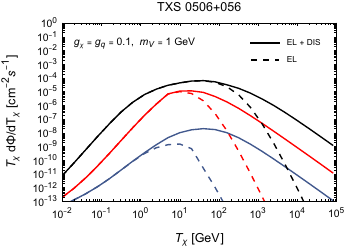}\\
\vspace{0.3cm}
\includegraphics[width=0.45\textwidth]{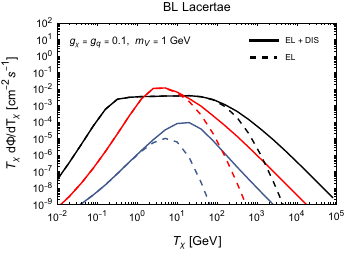}
\caption{The expected flux of BBDM from TXS 0506+056 (top panel) and BL Lacertae (bottom panel). The dashed lines represent the contributions from pure elastic scattering, while the solid lines also include the contribution from DIS. Different colours correspond to different DM mass $m_\chi$, namely 1 MeV (black), 10 MeV (red), and 100 MeV (blue).}
\label{fig:DM_Tx_Spectrum}
\end{figure}

In Fig.~\ref{fig:DM_Tx_Spectrum}, we present the results of BBDM flux for the blazars TXS 0506+056 (top panel) and BL Lacertae (bottom panel). The solid and dashed lines correspond to the full calculation (including both elastic and inelastic scattering) and the pure elastic scattering scenario, respectively. For each case, three DM masses are considered: \( m_\chi = 1 \) MeV (black), 10 MeV (red), and 100 MeV (blue). All results are derived under the assumption of \( g_\chi = g_q = 0.1 \) and \( m_V = 1 \) GeV.
From the figure, it is evident that in the low-energy region, the elastic scattering contribution dominates. In contrast, at higher energies, the DIS contribution becomes significant, allowing the BBDM flux to extend to much higher energy regimes. This result underscores the importance of including inelastic scattering effects in the analysis, as they play a crucial role in providing a more accurate description of the BBDM flux.

%%%%%%%%%%%%%%%%%%%%
\section{High energy neutrino and gamma ray flux from inelastic scattering}
\label{sec:gammav}
\begin{figure}
\centering
    \includegraphics[width=0.45\textwidth]{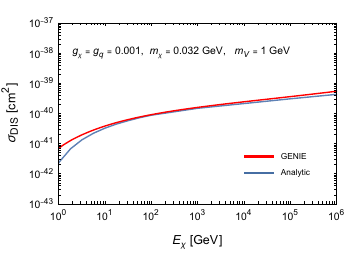}
    \caption{Comparison between the analytical computation of the DIS cross section (see Eq. \eqref{eq:ddxsec_dis}--\eqref{eq:disQ2}) and the numerical results given by \texttt{GENIE}. The input parameters are set as: $g_\chi = g_q = 0.001$, $m_\chi = 0.032$ GeV, $m_V = 1$ GeV.}
    \label{fig:geniecheck}
\end{figure}

As discussed in Sec.~\ref{sec:intro}, a key feature of including inelastic scattering processes is the production of high-energy neutrinos and gamma rays. In this work, we employ a specially modified version of \texttt{GENIE} to calculate the DIS process between BBDM and protons, which can be expressed as \cite{Andreopoulos:2015wxa,Berger:2018urf}. 
\begin{equation}
    \chi + p \to \chi + \text{mesons}(\text{e.g.}~\pi^\pm,\pi^0, ...) +\text{baryons}.
\end{equation}
In principle, the decay or deexcitation of all unstable mesons or baryons could produce neutrinos and/or photons. However, as the lightest mesons, $\pi$ mesons dominate the fragmentation process, with production rates significantly exceeding those of heavier mesons (e.g., $K$ mesons) and baryons (e.g., $\Delta$ resonances) \cite{ParticleDataGroup:2024cfk}. For simplicity, we consider only the contributions from $\pi$ mesons, with the corresponding decay chains given as follows:
\begin{equation}
    \pi^+ \to \mu^+ + \nu_\mu \to e^+ + \nu_e + \bar{\nu}_\mu + \nu_\mu,
    \label{eq:pi+}
\end{equation}
\begin{equation}
    \pi^- \to \mu^- + \bar{\nu}_\mu \to e^- + \bar{\nu}_e + \nu_\mu + \bar{\nu}_\mu,
    \label{eq:pi-}
\end{equation}
\begin{equation}
    \pi^0 \to \gamma \gamma.
    \label{eq:pi0}
\end{equation}
Note that we have neglected rare decay channels, such as $\pi^+ \to e^+ + \nu_e$ and $\pi^0 \to e^+ + e^- + \gamma$, and assumed a 100\% branching ratio for each decay step. After specifying the input parameters—particle masses ($m_\chi$, $m_V$), coupling constants ($g_q$, $g_\chi$), and the incoming DM energy $E_\chi$\footnote{Note that \texttt{GENIE} operates in the proton rest frame; transforming to the DM rest frame is straightforward via a Lorentz boost \cite{Berger:2018urf}.}—\texttt{GENIE} will calculate the DIS process, perform hadronization, and generate simulated events stored in a GHEP record \cite{Andreopoulos:2015wxa}. This record contains the four-momentum information of all final-state particles, including DM, mesons, and baryons. 
Our strategy is straightforward: we extract the $\pi$ mesons from the GHEP record, compute their decay dynamics (see Eqs.~\eqref{eq:pi+}--\eqref{eq:pi0}), and apply the appropriate Lorentz boosts to obtain the four-momenta of the resulting neutrinos and gamma rays. As a check, we compare the results of \texttt{GENIE} and analytic expressions for DIS of DM and protons (Fig. \ref{fig:geniecheck}). We randomly selected a set of parameters: $m_\chi = 0.032$ GeV, $m_V = 1$ GeV, $g_q=g_\chi = 0.001$. It can be seen that \texttt{GENIE} and the analytical results agree very well. \textcolor{black}{The discrepancy in the low-energy region arises from the internal cuts applied by \texttt{GENIE}, while our analytical calculation adopts a simplified threshold condition requiring the invariant mass of the final-state system (excluding the outgoing DM) to be larger than $ m_p + m_\pi$. This difference, however, does not impact our results, as the high-energy DIS scattering dominates the phenomenological implications discussed in this work.}

Finally, we achieve our goal: for incident protons of any energy in the DM rest frame, we obtain the four-momentum of all photons and neutrinos that are produced by $\pi$ mesons decay, and with these results, we can further derive the corresponding spectrum. Although the final-state photons and neutrinos are not collinear with the incident proton direction, we observe that the emission angle decreases rapidly as the proton energy increases. We define a critical angle $\theta_c$, which encloses 95\% of the emitted neutrinos or photons. This angle $\theta_c$ depends on the incident proton energy $E_p$. For simplicity, we assume an isotropic distribution of neutrinos or photons within a cone of half-opening angle $\theta_c$.

Taking neutrinos as an example, our simulation strategy proceeds as follows (the procedure for photons is analogous):
\begin{enumerate}
    \item Generate $N_p$ incoming protons with energy $E_p$, use \texttt{GENIE} to simulate the DIS events, and extract the four-momenta of all resulting neutrinos;\\
    \item Analyze the angular distribution of the emitted neutrinos and determine the critical angle $\theta_c^\nu$, which depends on the value of $E_p$. The corresponding solid angle is $\Delta\Omega = 2\pi (1 - \cos \theta_c^\nu)$;\\
    \item The normalized neutrino flux for each flavor is given by 
    \begin{equation}
        \frac{dN_\nu}{dE d\Omega}(E_p) = \frac{1}{3} \frac{dN_\nu}{dE}\frac{1}{N_p \Delta\Omega},
    \end{equation}
    where the factor $1/3$ accounts for neutrino oscillation effects over extragalactic distances;\\
    \item Integrate over the blazar proton spectrum (see Eq.~\eqref{eq:CRSpectrum}) to derive the final neutrino spectrum
    \begin{equation}
        \frac{dN_\nu}{dE d\Omega} = \int dE_p \int_0^{\theta_c^{\nu}(E_p)}  d\Omega\frac{d\Gamma}{dE_p d\Omega} \frac{dN_\nu}{dE d\Omega}(E_p),
    \end{equation}
    where $E_p = m_p + T_p$.
\end{enumerate}
Following the above steps, one can derive the neutrino and gamma ray spectrum around blazars. However, to facilitate meaningful comparisons with experimental data, both cosmological redshift and propagation effects must be carefully accounted for. Due to their extraordinarily weak interactions with matter, neutrinos are primarily affected by redshift effects alone. In contrast, photons are subject to significant propagation effects that cannot be neglected.

It is well established that gamma ray with energy exceeding approximately 100 GeV undergo substantial interactions with cosmic photons originating from stellar emissions, interstellar dust, and the relic radiation from the Big Bang \cite{Franceschini:2008tp,Planck:2018vyg}. Notably, pair production through interactions with the CMB and the extragalactic background light (EBL) renders the universe opaque to high-energy gamma rays traversing cosmological distances. The electrons and positrons generated in these interactions can subsequently upscatter photons from the same background radiation fields, producing secondary gamma rays. This cyclical process, known as an electromagnetic cascade, can significantly alter the spectral distribution of gamma rays over vast cosmological distances \cite{Murase:2012df,Berezinsky:2016feh,Venters_2010}.

\begin{figure}
\centering
    \includegraphics[width=0.45\textwidth]{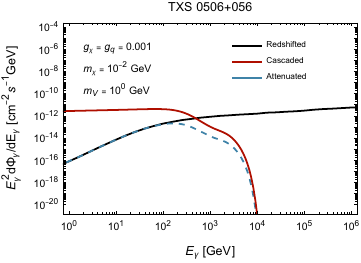}
    \caption{Propagation effects on high-energy gamma rays from TXS 0506+056. The black line shows the redshift effect only, the blue dashed line shows the attenuation effect only, and the red line represents the final observed gamma ray spectrum including cascading effects.}
    \label{fig:gammacas}
\end{figure}

In this work, we employ \texttt{$\gamma$-Cascade V4} to model the propagation effects of gamma rays \cite{Capanema:2024nwe}. The impact of these propagation effects on high-energy gamma rays originating from TXS 0506+056 is illustrated in Fig.~\ref{fig:gammacas}. The black line represents the scenario where only the redshift effect is considered, while the blue dashed line corresponds to the case where only the attenuation effect is accounted for. The red line depicts the final observed gamma ray spectrum, which includes the cascading effects. 
As shown in the figure, the propagation effects significantly alter the gamma ray spectrum, particularly in the energy region $E_\gamma \gtrsim \mathcal{O}(\text{TeV})$. Additionally, secondary emission substantially enhances the gamma ray flux below $\mathcal{O}(10^2)$ GeV, which facilitates detection.
\begin{figure}
\centering
\includegraphics[width=0.45\textwidth]{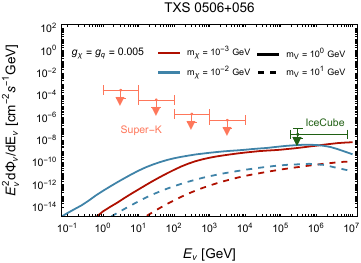}\\
\vspace{0.3cm}
\includegraphics[width=0.45\textwidth]{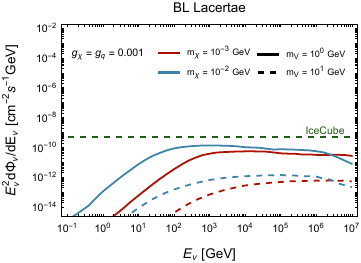}
\caption{The expected muon neutrino flux from TXS 0506+056 (top panel) and BL Lacertae (bottom panel). The solid lines represent the results of $m_V=1$ GeV, while the dashed lines are the results of $m_V=10$ GeV. Different colors correspond to different DM mass $m_\chi$, namely $10^{-3}$ GeV (red), $10^{-2}$ GeV (blue)\footnote{\textcolor{black}{At energies exceeding a few times $10^6 \, \text{GeV}$, the spectral shapes of the resulting neutrino fluxes exhibit a discernible dependence on the DM mass. Nonetheless, such behavior is anticipated to be largely mass-independent (see Fig. 1 in Ref.~\cite{DeMarchi:2024riu}). While a more refined numerical treatment may be requisite in this high-energy regime, it bears little relevance to the phenomenological focus of the present work.}}. The coupling constants are set as $g_q=g_\chi =0.005$ and $0.001$ for TXS 0506+056 and BL Lacertae, respectively. For comparison, we also demonstrate the observation or constraints from IceCube \cite{IceCube:2018dnn,IceCube:2019cia} and Super-K \cite{Super-Kamiokande:2019utr,Super-Kamiokande:2009uwx,Wang:2023zgk}
.}
\label{fig:v_Spectrum}
\end{figure}

\begin{figure}
\centering
\includegraphics[width=0.45\textwidth]{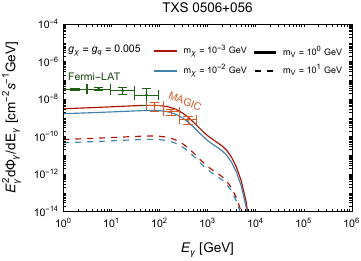}\\
\vspace{0.3cm}
\includegraphics[width=0.45\textwidth]{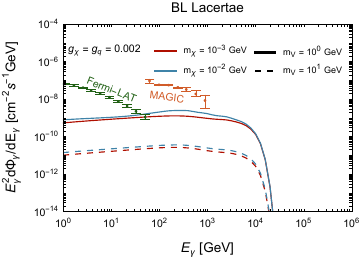}
\caption{The expected gamma ray flux from TXS 0506+056 (top panel) and BL Lacertae (bottom panel). The solid lines represent the results of $m_V=1$ GeV, while the dashed lines are the results of $m_V=10$ GeV. Different colors correspond to different DM mass $m_\chi$, namely $10^{-3}$ GeV (red), $10^{-2}$ GeV (blue). The coupling constant s are set as $g_q=g_\chi =0.005$ and $0.002$ for TXS 0506+056 and BL Lacertae, respectively.
For comparison, we also demonstrate the observation from Fermi-LAT \cite{Fermi-LAT:2015bhf,Mondal:2024gfn}, MAGIC \cite{Aleksi__2016,MAGIC:2019mfn}.}
\label{fig:gamma_Spectrum}
\end{figure}

The neutrino and gamma ray spectrum of TXS 0506+056 (top panel) and BL Lacertae (bottom panel), induced by the DIS process, are presented in Fig.~\ref{fig:v_Spectrum} and Fig.~\ref{fig:gamma_Spectrum}, respectively. The specific parameter selections are described in detail in the figure captions. For comparison, we also include the corresponding experimental observations and/or constraints from IceCube \cite{IceCube:2018dnn,IceCube:2019cia}, Super-K \cite{Super-Kamiokande:2019utr,Super-Kamiokande:2009uwx,Wang:2023zgk}, Fermi-LAT \cite{Fermi-LAT:2015bhf,Mondal:2024gfn}, and MAGIC \cite{Aleksi__2016}. In the next section, we will discuss the constraints derived from these experimental results.

%%%%%%%%%%%%%%%%%%%%
\section{Constraints on BBDM-Proton Scattering Cross Section}
\label{sec:constraints}
In this section, we derive constraints on the coupling constants between DM and protons by leveraging experimental data. Broadly speaking, these experiments can be categorized into two classes: ``direct'' and ``indirect.'' The former refers to experiments that probe the direct interactions between DM and detector targets, such as LZ and Borexino. The latter, on the other hand, relies on detecting neutrinos and photons produced as secondary signatures of DM interactions, as exemplified by Super-K, IceCube, Fermi-LAT, and MAGIC. Additionally, we incorporate the search for $Z$ boson decays at LEP to provide a comprehensive analysis.

To facilitate comparison with the literature, we define the DM-proton cross section $\sigma_{\chi p}$ as
\begin{equation}
    \sigma_{\chi p} = \frac{g_\chi^2 g_p^2}{\pi m_V^4} \mu_{\chi p}^2,
\end{equation}
where $\mu_{\chi p} = \frac{m_\chi m_p}{m_\chi + m_p}$ is the reduced mass. All experimental constraints will be translated onto the $m_\chi$-$\sigma_{\chi p}$ plane for analysis.

\subsection{The Z boson decay constraints from LEP}
\label{sec:LEP}
If $m_V<m_Z$, then the $Z$ boson can decay to $V_\mu$ and $\gamma$ through the quark loops, and the corresponding to a branching ratio is \cite{Dror:2017ehi}
\begin{equation}
    \frac{\Gamma_{Z\to\gamma V}}{\Gamma_Z} \simeq 6.8 \times 10^{-2} \left(\frac{\text{GeV}}{m_V}\right)^2 g_q^2.
    \label{eq:gammaV}
\end{equation}
In our model, the $V_\mu$ could further decay to $\chi$ (invisible state) or SM quarks $q$ (hadronic state), and the decay width can be expressed as
\begin{equation}
    \Gamma_{V\to f \bar{f}}=\frac{N_c g_f^2}{12 \pi}m_V\left(1+\frac{2m_f^2}{m_V^2}\right)\sqrt{1-\frac{4m_f^2}{m_V^2}},
    \label{eq:Vdecay}
\end{equation}
where $f\in \{q, \chi\}$, the color factor $N_c$ is 3 and 1 for $q$ and $\chi$, respectively. 
If $V_\mu$ decays invisibly, then the LEP search for a single photons at the $Z$ resonance limits this branching ratio to $\lesssim 10^{-6}$ \cite{L3:1997exg,DELPHI:1996qcc}, while bounds for hadronic decays of $V_\mu$ are less stringent, i.e. $\lesssim 3.6 \times 10^{-3}$ \cite{L3:1992kcg,OPAL:1991acn,L3:1991kow}. These experimental measures can be expressed as
\begin{align}
    \frac{\Gamma_{Z\to\gamma V}}{\Gamma_Z} \times \text{BR}(V\to \chi \bar{\chi}) &\lesssim 10^{-6}, \label{eq:Vinv}\\
    \frac{\Gamma_{Z\to\gamma V}}{\Gamma_Z} \times \text{BR}(V\to q \bar{q}) &\lesssim 3.6 \times 10^{-3}. \label{eq:Vhar}
\end{align}
It is important to note that the number of accessible quark flavors in the $V_\mu$ boson decays depends critically on its mass $m_V$. By combining the results from Eqs.~\eqref{eq:Vdecay}--\eqref{eq:Vhar}, we obtain constraints on the DM-proton scattering cross section $\sigma_{\chi p}$ derived from LEP searches, as indicated by the gray shaded region in Fig.~\ref{fig:DDLZ}. In this figure, the solid and dashed curves represent the cases for $m_V = 10$~GeV and $1$~GeV, respectively.  A noteworthy observation is the insensitivity of the LEP constraints to variations in $m_V$. The reason is as follows, for $g_q \gtrsim g_\chi$, the $\text{BR}(V\to q \bar{q})\sim 1$, which means $g_q/m_V$ is almost constant (see Eq. \eqref{eq:gammaV}). Moreover, along the exclusion boundary, the branching ratio to DM $\mathcal{B}(V\to\chi\bar{\chi}) \propto g_\chi^2/g_q^2$ also maintains a near constant value. Consequently, the exclusion lines of $\sigma_{\chi p}$ are independent of $m_V$ and proportional to the DM mass square, that is, $\sigma_{\chi p} \propto m_{\chi}^2$, which perfectly agrees with the observed scaling behavior of the exclusion limits.

\subsection{Direct detection from LZ and Borexino experiment}
\label{sec:LZ}
Direct detection experiments have emerged as a particularly promising approach for probing boosted DM \cite{PandaX:2024pme,CDEX:2024qzq,CDEX:2022fig,DeRomeri:2023ytt,Jho:2020sku}. Building upon the framework established in Ref. \cite{wang2021direct}, we perform refined calculations employing an improved BBDM flux that incorporates both energy-dependent effects and DIS contributions. Recently, the LZ experiment reported their latest DM search results using a total exposure of $4.2 \pm 0.1$ tonnes-year \cite{LZ:2024zvo}. For our analysis, we focus specifically on the WS2024 dataset, which accounts for $3.3$ tonne-years of exposure and provides the dominant contribution to the DM-nucleon exclusion limits. After comprehensive data selection procedures and salt removal, the expected background yields $1210 \pm 91$ events, compared to the observed count of $1220$ events \cite{LZ:2024zvo}. 
Applying the standard Poisson likelihood method \cite{Zyla:2020zbs}, we establish 90\% confidence level (C.L.) upper limits on the potential BBDM contribution, obtaining $N_\text{BBDM} = 55.99$ events. 

\begin{figure}
\centering
\includegraphics[width=0.45\textwidth]{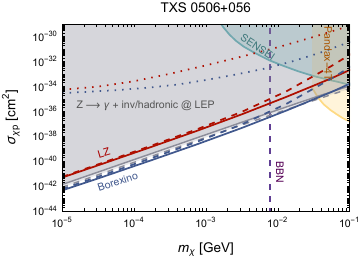}\\
\vspace{0.3cm}
\includegraphics[width=0.45\textwidth]{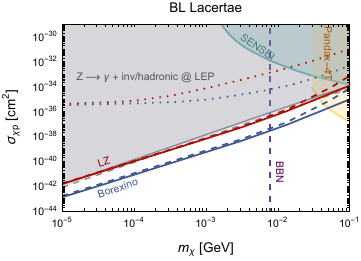}
\caption{The constraints on $\sigma_{\chi p}$ imposed by LZ (red) \cite{LZ:2024zvo} and Borexino (blue) \cite{Borexino:2013bot}. The top panel is for TXS 0506+056, while the bottom is for BL Lacertae. The solid and dashed lines correspond to $m_V=10$ GeV and $m_V=1$ GeV, respectively. The dotted line is for constant and elastic scattering scenarios for comparison \cite{wang2021direct}.  The constraints from Pandax-4T \cite{PandaX:2023xgl}, SENSEI \cite{SENSEI:2023zdf}, and BBN \cite{Giovanetti:2021izc} are included. The grey shaded area is excluded by LEP searches for $Z\to \gamma V$ \cite{L3:1997exg,DELPHI:1996qcc,L3:1992kcg,OPAL:1991acn,L3:1991kow}. }
\label{fig:DDLZ}
\end{figure}

The expected number of nuclear recoil events induced by BBDM interactions in the LZ detector can be expressed as:
\begin{equation}\label{eq:DM-N rate}
    N_\text{Xe}^\text{DM} \simeq  N_\text{Xe} t_\text{obs}\int_{T_\text{exp}^\text{min}}^{T_\text{exp}^\text{max}}dT_\text{Xe}\, \int_{T_{\chi}^\min(T_\text{Xe})}^{+\infty}\!\!\ 
  \frac{ d\sigma_\text{el}}{T_\text{Xe}}\frac{d\Phi_{\chi}}{dT_{\chi}}dT_\chi\,, 
\end{equation}
with
\begin{equation}
    \frac{d\sigma_{\rm el}}{dT_\text{Xe}} = 2 m_\text{Xe} \frac{ g_A^2g_\chi^2}{4\pi \beta^2} \frac{F^2(Q^2)}{(Q^2+m_V^2)^2} \left[ 1 -   \frac{\beta^2 Q^2}{Q^2_{\rm max}} + \frac{Q^4}{8 m_\text{Xe}^2 E_\chi^2} \right] \;,  \label{eq:DMXe_elas} 
\end{equation}
where $N_\text{Xe} = 2.53\times 10^{28}$ is the total number of xenon \cite{LZ:2024zvo}, $t_\text{obs} = 220$ days is the exposure time, 
$m_\text{Xe}$ is xenon mass,  $g_A = 132 g_p$ is the coupling constant between DM and xenon, 
$F(Q^2)$ is the conventional Helm form factor \cite{PhysRev.104.1466,Lewin:1995rx},
$\left[T_\text{exp}^\min,~T_\text{exp}^\max\right] = \left[5.4~\text{keV},~55~\text{keV}\right]$ is energy range of sensitivity of
LZ, $d\Phi_{\chi}/dT_{\chi}$ is the BBDM flux at detector. The corresponding constraints on $\sigma_{\chi p}$ from TXS 0506+056 and BL Lacertae can be obtained by imposing 
\begin{equation}
    N_\text{Xe}^\text{DM} < N_\text{BBDM}.
    \label{eq:limit}
\end{equation}
The derived constraints on $\sigma_{\chi p}$ from TXS 0506+056 and BL Lacertae are presented in the upper and lower panels of Fig.~\ref{fig:DDLZ}, respectively. Our full analysis results, shown by the red solid ($m_V=10$ GeV) and dashed ($m_V=1$ GeV) curves, demonstrate significant improvement over the simplified constant cross section and elastic-only scenarios (red dotted lines) previously reported in \cite{wang2021direct}. This comparison underscores the importance of our complete treatment incorporating both energy-dependent effects and inelastic scattering contributions. For context, we also display complementary bounds from SENSEI \cite{SENSEI:2023zdf} and PandaX-4T \cite{PandaX:2023xgl}.

Beyond conventional direct detection methods, neutrino experiments such as Borexino can serve as powerful complementary detectors for DM searches \cite{Borexino:2013bot}. Compared with the LZ experiment, Borexino has higher thresholds but significantly larger target mass, which provides additional sensitivity to certain regions of the parameter space. 
Following the analysis in Refs. \cite{Bringmann:2018cvk,Bell:2023sdq,wang2021direct}, the limiting scattering rate per proton can be derived from Borexino, 
that is 
\begin{equation}
 \Gamma_p(T_p > 25 ~\text{MeV}) < 2 \times 10^{-39}~\text{s}^{-1},
\end{equation}
where we have used the approximation that the ratio between quenched energy deposit (equivalent electron energy $T_e$) and proton recoil energy $T_p$ fulfills $T_e(T_p)/T_p \approx 2$ for $T_p \gtrsim 5$ MeV \cite{Beacom:2002hs,Dasgupta:2011wg}. Similarly to LZ results, the constraints on $\sigma_{\chi p}$ from Borexino are shown in blue curves in Fig. \ref{fig:DDLZ}. Due to the larger exposure, Borexino gives stronger constraints than LZ. This demonstrates the crucial complementarity between conventional direct detection and neutrino experiments in probing DM parameter space.

\subsection{Indirect detection from Super-K, IceCube, Fermi-LAT, and MAGIC experiment}
\label{sec:icecube}

As emphasized in Section~\ref{sec:gammav}, the DIS process generates high-energy gamma ray and neutrino fluxes that are detectable by both ground- and space-based observatories. This provides a complementary ``indirect" detection channel for BBDM, exhibiting distinct spectral features that differentiate it from constant cross section or pure elastic scattering scenarios. 
For neutrino detection, we analyze data from Super-K \cite{Super-Kamiokande:2019utr,Super-Kamiokande:2009uwx,Wang:2023zgk} and IceCube \cite{IceCube:2018dnn,IceCube:2019cia}, while gamma ray observations are drawn from Fermi-LAT \cite{Fermi-LAT:2015bhf,Mondal:2024gfn} and MAGIC \cite{Aleksi__2016,MAGIC:2019mfn}. By combining direct observational data and point source constraints from these experiments, we derive robust limits on $\sigma_{\chi p}$. Our methodology is conceptually straightforward: we require that the predicted neutrino or gamma ray fluxes do not exceed the experimental measurements or upper limits.

\noindent \textit{\textbf{Super-K}}---The Super-K collaboration has conducted a comprehensive search for astrophysical neutrinos in the GeV--TeV energy range, particularly in the direction of the blazar TXS 0506+056. Their 90\% confidence level upper limit on the $\nu_\mu + \bar{\nu}_\mu$ flux is represented by the orange data points in Fig.~\ref{fig:v_Spectrum} \cite{Super-Kamiokande:2019utr}. For BL Lacertae, we utilize Super-K's extragalactic point source search results \cite{Super-Kamiokande:2009uwx,Wang:2023zgk}, which provide constraints on the integrated flux $\Phi(E_\nu > 1.6~\text{GeV})$ as a function of declination. The BL Lacertae's declination is $42^\circ 16' 40''$, which corresponds to:
\begin{equation}
    \Phi(E_\nu > 1.6~\text{GeV}) < 1.87 \times 10^{-7}~\text{cm}^{-2}~\text{s}^{-1}.
\end{equation}
We then compare this limit with the energy-integrated neutrino spectrum predicted for BL Lacertae.

\noindent \textit{\textbf{IceCube}}---The IceCube detection of the high-energy neutrino event IceCube-170922A, spatially coincident with TXS 0506+056, provides an important upper limit on the muon neutrino flux (green data point in Fig.~\ref{fig:v_Spectrum}). For conservative analysis, we adopt the flux limit derived from 0.5 years of observation \cite{IceCube:2018dnn}. For BL Lacertae, we utilize results from IceCube's decade-long (2008--2018) point source search \cite{IceCube:2019cia}, which constrains the muon neutrino flux to $4.9 \times 10^{-10}~\text{GeV}^{-1}~\text{cm}^{-2}~\text{s}^{-1}$ (green dashed line in Fig.~\ref{fig:v_Spectrum}).

\noindent \textit{\textbf{Fermi-LAT and MAGIC}}---Both Fermi-LAT and MAGIC have obtained high-quality spectral measurements of the gamma ray emission from TXS 0506+056 and BL Lacertae. The observational data are shown as green (Fermi-LAT) and orange (MAGIC) points in our spectral plots.

\begin{figure}
\centering
\includegraphics[width=0.45\textwidth]{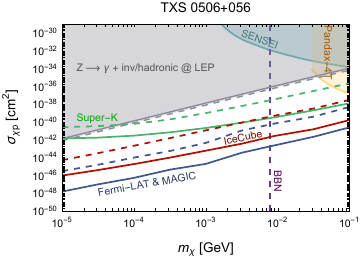}\\
\vspace{0.3cm}
\includegraphics[width=0.45\textwidth]{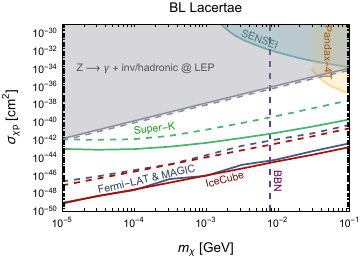}
\caption{The constraints on $\sigma_{\chi p}$ imposed by Super-K (green) \cite{Super-Kamiokande:2019utr,Super-Kamiokande:2009uwx,Wang:2023zgk}, IceCube (red) \cite{IceCube:2018dnn,IceCube:2019cia}, Fermi-LAT/MAGIC (blue) \cite{Fermi-LAT:2015bhf,Mondal:2024gfn,Aleksi__2016,MAGIC:2019mfn}. The top panel is for TXS 0506+056, while the bottom is for BL Lacertae. The solid and dashed red lines correspond to $m_V=10$ GeV and $m_V=1$ GeV, respectively. The constraints from Pandax-4T \cite{PandaX:2023xgl}, SENSEI \cite{SENSEI:2023zdf}, and BBN \cite{Giovanetti:2021izc} are included. The grey shaded area is excluded by LEP searches for $Z\to \gamma V$ \cite{L3:1997exg,DELPHI:1996qcc,L3:1992kcg,OPAL:1991acn,L3:1991kow}. }
\label{fig:IDexp}
\end{figure}

In Fig.~\ref{fig:IDexp} we present the combined constraints from Super-K (green), Fermi-LAT/MAGIC (blue), and IceCube (red), with solid and dashed curves corresponding to $m_V = 10$~GeV and 1~GeV, respectively. Notably, the indirect detection constraints on $\sigma_{\chi p}$ are several orders of magnitude stronger than those obtained from direct detection experiments, highlighting the exceptional sensitivity of astrophysical observations to boosted DM signatures.

%%%%%%%%%%%%%%%%%%%%%%%%%%%%
\section{Conclusion}
\label{sec:Conclusion}
Due to the extremely high powerful jet and the large DM density in the central region, blazars are  ideal DM boosters.
Following the framework established in Refs.~\cite{wang2021direct,Granelli:2022ysi}, we focus on two prominent blazars—TXS 0506+056 and BL Lacertae—within a specific model where DM interacts with SM quarks via a massive vector mediator $V_\mu$. Our calculations of BBDM flux incorporate both energy-dependent effects and contributions from DIS process, revealing that the DIS could extend the BBDM spectrum to significantly higher energies (see Fig.~\ref{fig:DM_Tx_Spectrum}).

Consistent with previous studies \cite{wang2021direct,Granelli:2022ysi}, we demonstrate that BBDM can produce detectable signals in terrestrial detectors. In this work, we utilize LZ and Borexino as effective ``direct" DM detectors to constrain $\sigma_{\chi p}$. Our analysis shows that, compared to the constant cross section and pure elastic scattering scenarios, these constraints improve by several orders of magnitude, reaching sensitivity comparable to LEP's $Z\to \gamma V$ search limits. This striking enhancement underscores the critical importance of including full energy-dependent and inelastic effects in the theoretical treatment.

Furthermore, the DIS process between protons and DM generates substantial pion production, with subsequent decays yielding high-energy neutrinos and gamma rays. These secondary particles propagate to Earth, where they can be detected by neutrino (Super-K, IceCube) and gamma ray (Fermi-LAT, MAGIC) observatories. By comparing our predictions with observational data or upper limits from these experiments—while properly accounting for neutrino oscillation, cosmological redshift, and propagation effects—we derive stringent constraints on $\sigma_{\chi p}$. Remarkably, these ``indirect" detection methods yield limits that are significantly stronger than those obtained from direct detection experiments, representing a substantial advance beyond previous work \cite{wang2021direct,Granelli:2022ysi}.

Of particular interest is the realization that DM-proton DIS processes constitute a novel source of high-energy astrophysical neutrinos. For energetic sources like blazars, this neutrino production channel becomes especially important. The connection to IceCube's 2017 observation of high-energy neutrinos from the TXS 0506+05 direction suggests exciting possibilities: the high-energy neutrino event may come from DM! This interpretation motivates a comprehensive multi-messenger approach that synergistically combines neutrino observations, gamma ray spectrum, and direct DM detection signatures. Such combined analysis offers unprecedented potential to constrain blazar emission models and probe the particle nature of DM, potentially resolving long-standing questions in both high-energy astrophysics and particle cosmology.

\textcolor{black}{It should be noted that several additional factors may also impact our results. For instance, the complex environment at the center of blazars may lead to absorption of high-energy gamma rays \cite{MAGIC:2018sak}; the actual geometry of the jet may deviate from the idealized conical structure assumed in this work \cite{Cui:2024ggx,Mizuno:2007fd}; the trajectories of protons may be influenced by strong magnetic fields near the central engine, deviating from straight-line motion \cite{EventHorizonTelescope:2021srq}; and the precise launching point of the jet remains uncertain \cite{Nakamura:2018htq}. A thorough investigation of these effects is beyond the scope of this paper and will be pursued in future work.}\\

\textcolor{black}{Note added: During the preparation of our manuscript, we became aware of a closely related study \cite{DeMarchi:2024riu}, in which the authors explored the neutrino signals induced by inelastic scattering between DM and protons, focusing on two specific blazar sources: TXS 0506+056 and AP Librae. Compared to that work, our study presents a more comprehensive phenomenological investigation by incorporating not only neutrino signals but also constraints from direct detection experiments and high-energy photon observations. Together, these two analyses provide complementary insights, enriching the broader understanding of the potential astrophysical signatures associated with the BBDM.}

%%%%%%%%%%%%%%%%%%%%%%%%%%%%
\begin{acknowledgments}
The authors wish to thank Qinrui Liu, Alessandro Granelli, Chen Xia, Su-Jie Lin, and Zhao-Huan Yu for their thoughtful discussions and suggestions. This work was supported by the National Natural Science Foundation of China (NSFC) under Grants
No. 12405119 and the Natural Science Foundation of Sichuan Province under Grant No. 2025ZNSFSC0880.
\end{acknowledgments}

\newpage
\bibliography{BBDMSK}

\end{document}